\documentclass[a4paper,fleqn, authoryear]{cas-dc}

\usepackage[round, authoryear, sort]{natbib}
\usepackage{soul}
\usepackage{amsmath}
\usepackage{graphicx}
\usepackage{pifont}
\usepackage{caption}
\definecolor{RED}{RGB}{156,78,90}

\def\tsc#1{\csdef{#1}{\textsc{\lowercase{#1}}\xspace}}
\tsc{WGM}
\tsc{QE}
\tsc{EP}
\tsc{PMS}
\tsc{BEC}
\tsc{DE}

\begin{document}
\captionsetup[figure]{labelfont={bf},labelformat={default},name={Fig.},labelsep=period}


\let\WriteBookmarks\relax
\def\floatpagepagefraction{1}
\def\textpagefraction{.001}

\shorttitle{Biphasic heat transfer enhancement}
\shortauthors{H. Qi et~al.}

\title [mode = title]{Active biphasic heat transfer enhancement in vertical natural convection} 


\author[1]{Haoran Qi}
\credit{Methodology, Validation, Investigation, Writing original draft, Writing-review $\&$ editing, Formal analysis, Data curation, Visualization}

\author[1]{Chaoben Zhao}
\credit{Methodology, Software, Validation, Investigation, Writing original draft, Visualization}

\author[1]{Yaning Fan}
\credit{Validation, Writing-Review \& Editing, Project administration}

\author[1]{Yihong Du}
\credit{Validation, Writing-Review \& Editing, Project administration}

\author[2,3]{Varghese Mathai}
\cormark[1]
\ead{vmathai@umass.edu}
\cortext[cor1]{Corresponding author}
\credit{Supervision, Writing-Review \& Editing}

\author[1,4]{Chao Sun}
\cormark[2]
\cortext[cor2]{Corresponding author}
\ead{chaosun@tsinghua.edu.cn}
\credit{Supervision, Resources, Funding acquisition, Project administration, Writing-Review \& Editing}

\address[1]{New Cornerstone Science Laboratory, Center for Combustion Energy, Key Laboratory for Thermal Science and Power Engineering of Ministry of Education, Department of Energy and Power Engineering, Tsinghua University, 100084 Beijing, China}
\address[2]{Department of Physics, University of Massachusetts Amherst, Massachusetts 01003, USA}
\address[3]{Department of Mechanical \& Industrial Engineering, University of Massachusetts Amherst, Massachusetts 01003, USA}
\address[4]{Department of Engineering Mechanics, School of Aerospace Engineering, Tsinghua University, Beijing 100084, China}

\begin{abstract}
Vertical natural convection (VC), often cannot meet the high heat transfer demands due to the inherent misalignment of the direction of buoyancy (vertical) with the direction of the heat transfer (horizontal). Here we applied a novel strategy on a water based VC system to enhance the heat transfer. By adding {2\% of the total volume with a low-boiling-temperature liquid (HFE-7000) and introducing a gas-liquid layer on top of the VC cell, we create a self-sustained state of pseudo-turbulence with evaporating, circulating and condensing biphasic bubbles}. The system achieves 246\% heat transfer enhancement at constant superheat $T_{sup}\approx6.4~\text{K}$ when the liquid in the full nucleate boiling state. Using shadowgraphy and Laser Doppler Anemometry (LDA) methods, {we validate that the bubbles and biphasic particles induced agitation enhances the heat flux and modifies the temperature field of the heat transfer.}

\end{abstract}


\maketitle
\section{Introduction}

Natural convection, driven by buoyancy, is found in natural and industrial contexts such as stellar interiors, atmospheric/ocean convection, HVAC systems, electronic cooling, and nuclear reactors. A key geometry for studying natural convection is the Rayleigh-Bénard convection (RBC) cell, featuring a heated bottom, a cooled top, and fluid in between. Heat transfer in natural convection follows classic scaling laws \citep{lohse2010small}.
Natural convective heat exchange can be restricted if buoyancy does not align with the desired direction of heat transfer. Vertical convection (VC) \citep{batchelor1954heat,patterson1980unsteady,paolucci1990direct,shishkina2016momentum,wang2021regime} serves as a primary example, where fluid is heated on one side (a vertical wall) and cooled on the other \citep{thorpe1969effect,tanny1988dynamics}. A key difference from RBC is that in VC, gravity is not aligned with the intended heat flux, resulting in a weaker effective heat transfer rate. Yet, VC is widely encountered in many energy transfer systems, including building ventilation, thermal power plants, and in electronic cooling applications. Enhancing heat transport in this VC is, therefore, of great relevance in a variety of applications.

\begin{figure*}
\centering
\includegraphics[width=0.65\linewidth]{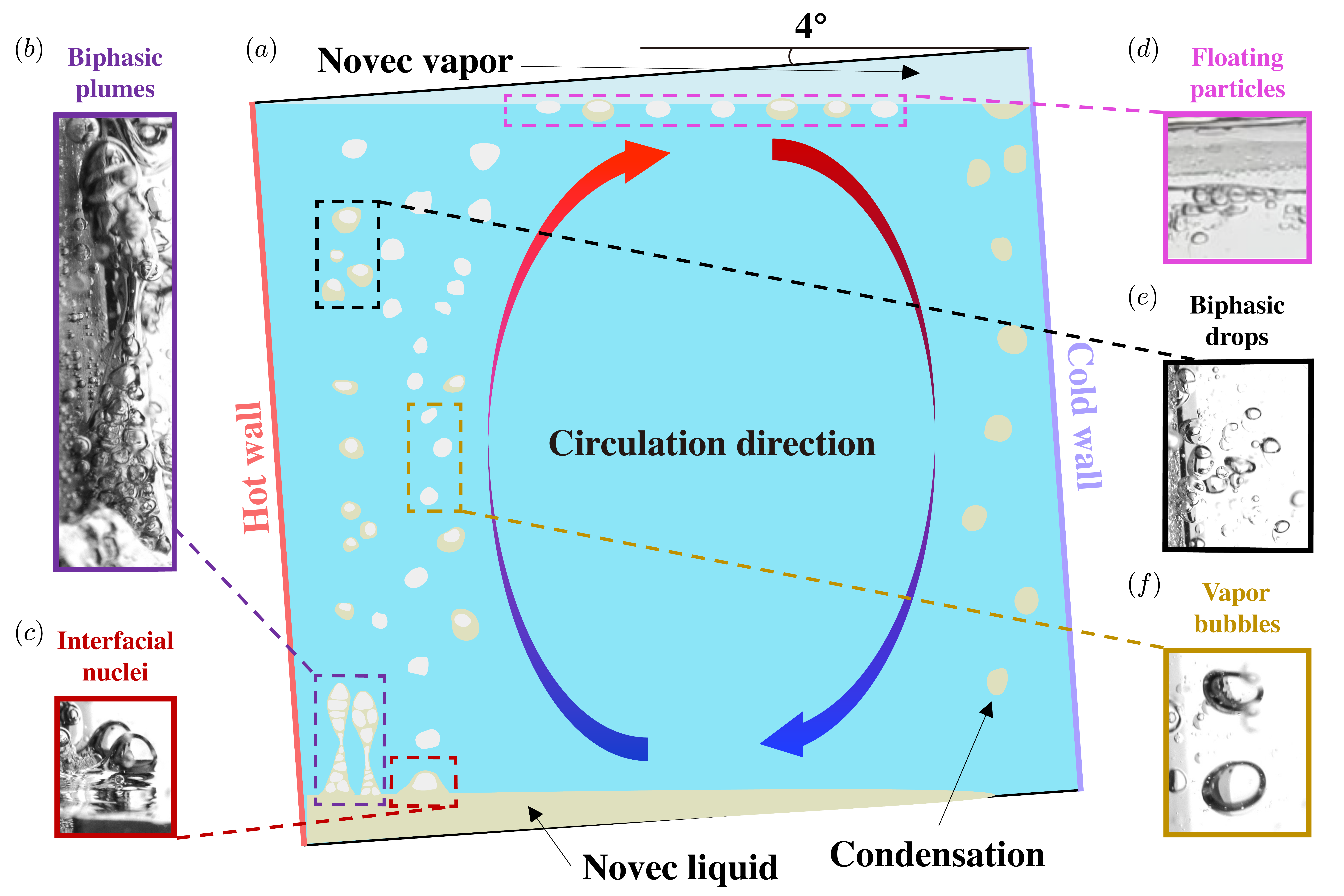}
\caption{Schematic and biphasic organizations of the biphasic VC. ($a$) shows the schematic of the setup when HFE is fully nucleate boiling. ($b$), ($c$), ($d$), ($e$), and ($f$) are the photographs of the nucleate boiling-induced organizations whose counterparts in ($a$) are framed by dashed outlines of the same color. Note that biphasic organizations in ($b$) and ($e$) only exist in a fully nucleate boiling regime.}
\label{fig:bubbles}
\end{figure*}

Some of the earliest works on VC were conducted in the context of steady-state heat transfer in windows \citep{batchelor1954heat}. Interest in VC has grown, particularly to understand the flow state transitions. In contrast to RBC, VC is characterized by the stabilizing nature of the stratified flow \citep{de1983natural}. Hence the transitions are more gradual in VC, and as the Rayleigh number ($Ra$) is increased, {periodic, quasi-periodic \citep{janssen1995influence}, or chaotic flows emerge \citep{paolucci1989transition,le1998onset}. At high $Ra$, the VC flow does indeed transition to a turbulent state \citep{paolucci1990direct,wang2021regime}.} 
Research on VC has increasingly focused on its heat transfer properties, particularly the scaling relationships among the Nusselt number ($Nu$), Reynolds number ($Re$), and control parameters $Ra$ and $Pr$ \citep{howland2022boundary,ke2020law}.

In recent years, there has been a growing interest in strategies to enhance heat transport in VC systems. By spatially harmonic temperature modulation and rough walls, \cite{chong2024heat} and \cite{meng2024heat} create coherent thermal plumes, thereby enhancing thermal turbulence.
However, the maximum enhancement was about 190\% \citep{jiang2019convective} and moreover, it was required to increase the temperature difference by an order of magnitude, to $Ra \approx 10^{11}$. Vibrating walls can alter flow structures and heat transport. \cite{guo2022pof} studied vertical vibration in a square convection cell, revealing that oscillatory effects enhance heat transfer efficiency by up to 600\% by inducing laminar-turbulent transitions, thereby increasing the scaling exponents of the vibrational Rayleigh number. Heat transfer in an open cell VC can be enhanced dramatically, by up to 2000\%, by injecting gas bubbles \citep{kitagawa2008heat,gvozdic2018experimental,gvozdic2019experimental}. However, all of these strategies required either external energy input or major modifications to the existing configuration of the convection cell.

\cite{wang2019self,wang2020experimental} showed that a biphasic state of natural convection can be triggered by introducing a small fraction of HFE-7000 into a water-based Rayleigh-B\'enard convection system and access 800\% heat transfer enhancement. 
Here we explore the question of whether a self-sustained biphasic state can be realized in VC, even though the direction of buoyancy and heat transfer are misaligned. 
We developed an experimental strategy utilizing biphasic boiling of a thin HFE-7000 layer at the bottom, combined with a gas-liquid layer facilitating horizontal transport of the floating bubbles to the colder side wall (see Fig.~\ref{fig:bubbles}. Here we have a brief depiction of the experimental methods in Section \ref{sec:method}, by which we measure non-monotonic heat transfer characteristics in Section \ref{sec:heat transfer}. We then verify the underlying mechanisms by shadowgraph images and LDA in Section \ref{sec:mechanism}, followed by a short conclusion (Section \ref{sec:Summary}).

\section{Methods}\label{sec:method}
 
We conducted the experiments in a rectangular convection cell {of 3500~ml volume capacity} with $4^{\circ}$ inclination to achieve the circulation of the condensing and boiling as shown in \textbf{Supplementary Materials}. By exerting isothermal and constant heat conditions on the cold and hot walls, respectively, we can control and keep the temperature of the two walls. Besides the power per area $q~\rm{(kW/m^{2})}$, we use its dimensionless parameter Nusselt number $Nu$, which is defined in \textbf{Supplementary Materials} with definition of another dimensionless parameter as well as the thermodynamic parameters of HFE-7000. The additive HFE-7000 who has a low boiling point 
$T_{sat}\approx34.2\rm~^{\circ}C$ at 1 atm is used to induce boiling at a safe temperature. We apply shadowgraphs on the flow (with 20\% glycerol
 in water) to visualize the thermal instabilities. To measure the liquid velocity of the working fluid, water, we apply Laser Doppler Anemometry (LDA) whose laser focuses are at the measurement position as shown in Fig.~\ref{fig:wall_velocity}$a$.

\section{Heat transfer enhancement}\label{sec:heat transfer}
\begin{figure}[!tbp]
 \centering
 \includegraphics[width=1\linewidth]{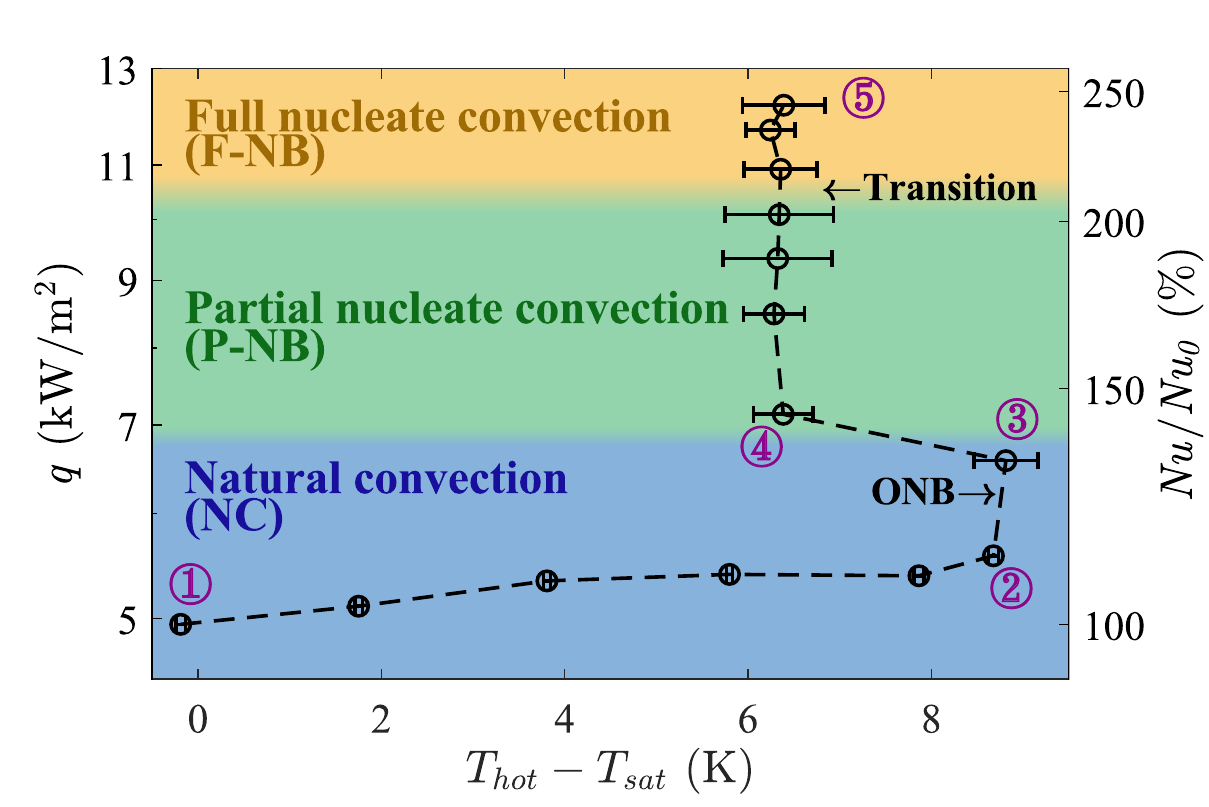}
  \caption{Wall heat flux density $q\rm~(kW/m^{2})$ and normalized Nusselt number $Nu/Nu_{0}$ v.s. wall superheat $T_{sup}=T_{hot}-T_{sat}$. Here, $Nu_{0}\approx103$ is the lowest Nusselt number in our experiments. The magenta circled numbers ~$1\ding{182}$~ and ~$5\ding{182}$~ are for the origin and the end of all the cases, and ~$2\ding{182}$~, ~$3\ding{182}$~, and ~$4\ding{182}$~ are for three cases transitioning from one state to another. Errorbars are based on the standard deviation of the time series of the Nusselt number.}
\label{fig:heat_transfer}
\end{figure}

Here, we first examine the boiling-induced collective dynamics of the bubbles and droplets, as well as the global heat transfer in VC. These vary significantly upon increasing the superheat, $T_{sup}$. Figure.~\ref{fig:heat_transfer} illustrates the relationship between heat flux $q$ (the left axis), the controlling parameter in experiments, as well as its dimensionless expression $Nu$ (the right axis) vs. superheat $T_{sup}=T_{\text{hot}} - T_{\text{sat}}$ of the bottom plate, the response parameter. Three distinct regimes are observed. At low heat flux levels (~$1\ding{182}-2\ding{182}~$), the system remains in the single-phase natural convection regime, for which the heat flux remains stable. As the power input to the hot wall is increased further, the system transitions (~$2\ding{182}-3\ding{182}~$) into the so-called partial nucleate boiling (P-NB) regime, marked by a marginal rise in wall superheat, as this coincides with a noticeable increase in the wall heat flux from 5.5 kW/cm$^2$ to 6.2 kW/cm$^2$. With a further increase of the input power, the system enters the full nucleate boiling (F-NB) regime, characterized by a significant increase in wall heat flux. Remarkably, the wall superheat decreases to 6.4 K, while the wall heat flux increases by 246$\%$. The three regimes are highlighted in blue, green, and orange shaded regions in Fig.~\ref{fig:heat_transfer} and will be hereafter referred to NC, P-NB, and F-NB, respectively. The biphasic activity as the VC system transitions from NC to P-NB and to F-NB is dictated by the appearance of nucleation sites of boiling on the HFE-water (liquid-liquid) interface at the bottom surface, the formation of biphasic plumes, the horizontal transport of floating bubbles at the gas-liquid interface at the top of the cell, and the sinking of biphasic droplets near the colder plate. {Note that all transitions from one regime to another are unsteady.} A schematic of the VC setup along with the biphasic activity is displayed in  Fig.~\ref{fig:bubbles}$a$-$f$.

As the side plate is heated above the boiling temperature of HFE-7000, a steady stream of tiny vapor bubbles forms at seemingly random nuclei on the surface (see Fig.\ref{fig:bubbles} $c$). The bubbles detach and coalesce into larger ones, increasing the effective buoyancy. Not until the bubble size reaches the millimetric scale is buoyancy strong enough to detach them. As shown in the blue shaded regime of Fig.~\ref{fig:heat_transfer}, until this regime the heat flux increases only slightly with wall superheat, with an $8.8 \rm~K$ temperature increment leading to merely a 10\% enhancement in heat transfer (~$1\ding{182}-3\ding{182}~$). This modest improvement can be attributed to the latent heat of phase change, which coincides with the significant increase in $T_{sup}$, as the Rayleigh number is held constant and single phase contribution to heat flux is not expected to increase either due to tilt or due to phase change \citep{guo2015effect,grossmann2001thermal,stevens2011prandtl}.

\begin{figure*}
 \centering
 \includegraphics[width=0.8\linewidth]{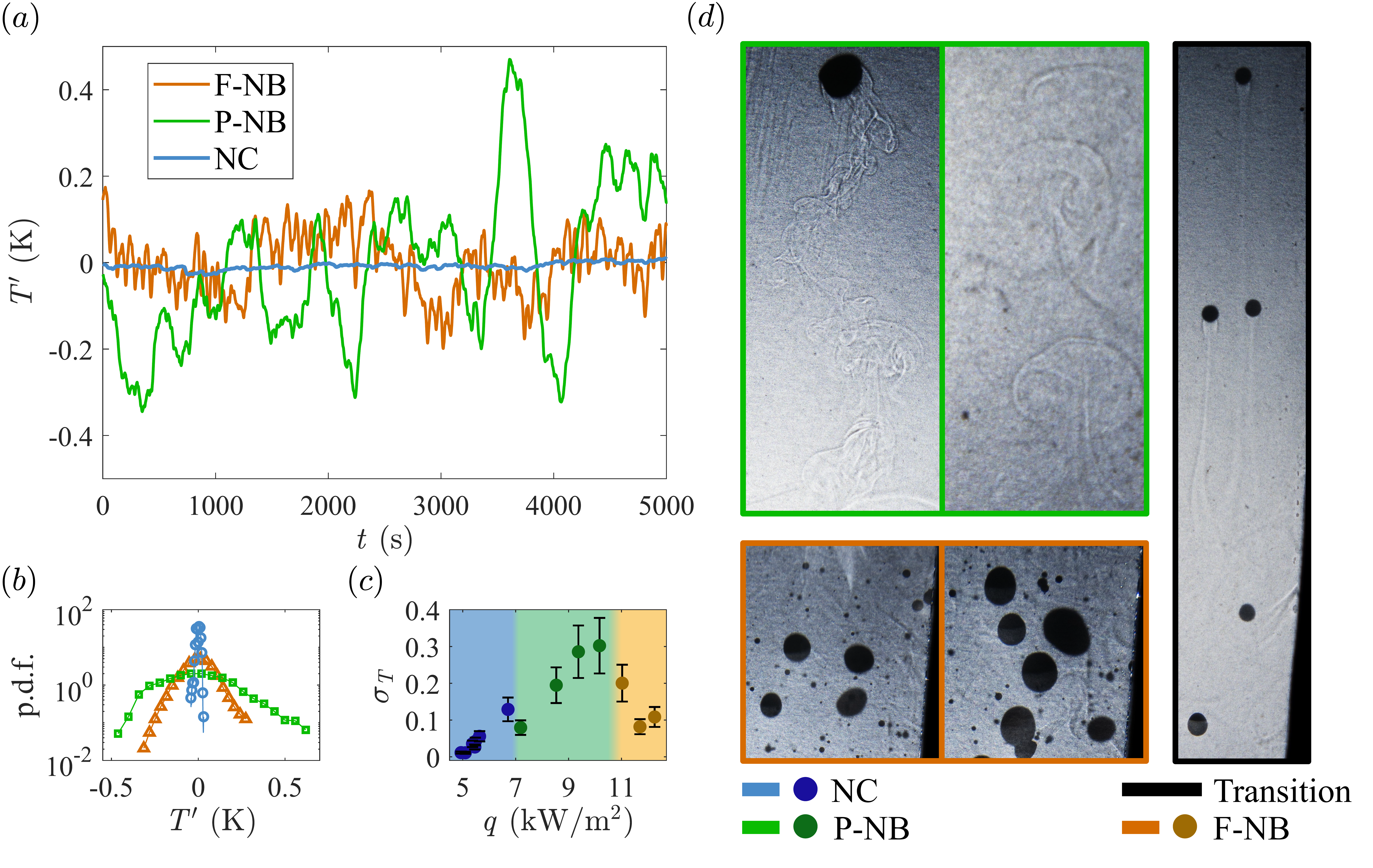}
  
\caption{Statistical temperature signals and shadowgraphs. ($a$) Temperature deviations time series for the hot wall in three regimes, NC, P-NB, and F-NB. ($b$) and ($c$) are the probability density function (p.d.f.) and standard deviation ($\sigma_{T}$) of the same signals. ($d$) is the shadowgraph images of bubbles and thermal organizations in P-NB, F-NB, and the transition between them, as marked by outlines in green, orange, and black. The organizations in the left green boxes are the first few wakes twisted by the bubble's trajectory, and the right is the far field wakes. The orange outlines show the thermal plumes rooted on the hot wall and the bubbles passing by without any thermal wakes. The black outlines enclose the upper three bubbles with thermal wakes but two lower bubbles without thermal wakes, indicating the flow is more homogeneous in temperature in the far field.}
\label{fig:thermal_characteristics}
\end{figure*}

When the superheat reaches $T_{sup}\approx8.8\rm~K$ and the power input keeps increasing, the effective buoyancy to viscous force on the bubbles is large enough to detach them. This yields a bubble Reynolds estimate of a few hundred, implying a turbulent wake. Although the wall superheat rises only slightly, a sharp increase in bubble generation rate is observed, resulting in noticeable biphasic activity. The increase in biphasic activity also coincides with a sudden drop in the wall temperature. When the power input is further increased, vapor bubbles are formed, as shown in Fig.~\ref{fig:bubbles}$f$. The system ultimately stabilizes at a reduced superheat $T_{sup}\approx6.4\rm~K$ and reaches the P-NB regime indicated by the green shading in Fig.~\ref{fig:heat_transfer}. The wall superheat becomes independent of wall heat flux and remains still at $T_{sup} \approx 6.4\rm~K$ (~$4\ding{182}-5\ding{182}~$). Within this regime, an increase in bubble generation correlates with an enhanced heat transfer. Once the convection enters the P-NB regime, the latent heat plays only a marginal role in the heat transfer. The wall heat flux is enhanced by up to $246\%$, and the Nusselt number correspondingly increases by $240\%$, {from which a discrepancy between two numbers caused by changes on property parameters due to increasing vapor volume}, as indicated in the orange shaded region in Fig.~\ref{fig:heat_transfer}. We halt further increase of the wall heat flux when the heat transfer reaches 271\% in $Nu/Nu_{0}$ because {of the uncontrollable boiling crisis under this insufficient cooling rate (13.6K below HFE-7000's boiling point)}.

In the F-NB regime, the bubble generation rate and detachment rate are significantly elevated compared to other regimes. Biphasic plumes comprised of multiple bubbles enveloped by liquid HFE are generated (see Fig.~\ref{fig:bubbles}$b$). These bubbles will be referred to as biphasic risers.  Finally, all the biphasic risers approach the gas-liquid interface at the top (see Fig.~\ref{fig:bubbles}$d$). The next phase of transport is enabled by the influence of the large-scale circulations in the cell. The floating biphasic bubbles glide along the almost frictionless gas-liquid interface, supported by the unique and essential design of the VC system. The drag exerted by the current on the floating bubbles enables this transport and is crucial for the self-sustained cycle in the VC system. As the floating bubbles approach the colder side plate, they begin to condense and sink. However, the boiling-transport-condensation cycle of the biphasic particles accounts for only 1.6\% of the total heat transfer (see \textbf{supplemental Materials}). Thus, the heat transfer enhancement is not dominated by the phase change.

\section{Mechanism on thermal characteristics}\label{sec:mechanism}
To assess the primary mechanism of heat transfer enhancement, we measure the temporal characteristics of liquid temperature during the NC, P-NB, and F-NB regimes. The signal showing the temperature series and its statistical properties effectively distinguishes the three regimes. The time series displays temperature deviation of the hot wall in each regime, as shown in Fig.~\ref{fig:thermal_characteristics}$a$, representing the $1^{st}$, $9^{th}$ and $13^{th}$ cases for which $q = 4.94\rm~kW/m^{2}$, $q = 8.53\rm~kW/m^{2}$, and $q = 11.7\rm~kW/m^{2}$, respectively, corresponding to NC, P-NB, and F-NB here and after, respectively. In the NC regime, the temperature deviations are small in amplitude. In contrast, the regime P-NB is characterized by partially nucleate boiling, and exhibits very strong fluctuations, Surprisingly, the regime with even larger heat flux, namely the F-NB, shows a reduction in the thermal fluctuation amplitude. This non-monotonic behavior is also reflected in the probability density functions (p.d.f.), where the p.d.f. in the NC regime has the narrowest range of thermal fluctuations, the P-NB regime exhibits a dramatic widening of the tails of the distribution, and the fluctuations again reduce to an intermediate level for F-NB (see Fig.~\ref{fig:thermal_characteristics}$b$). 

The standard deviation of temperature fluctuations is shown in Fig.~\ref{fig:thermal_characteristics}$c$. An abrupt increase in fluctuations, corresponding to P-NB, is attributed to phase change induced by the boiling-condensation cycle, which leads to extreme and intermittent ``cooling'' and ``heating" of different regions on the hot plate as boiling absorbs latent heat. The evidence that this increase is attributed to latent heat is evident from the fact that even though the fluctuations are increased, the frequency of the thermal fluctuations is low. The large amplitude temperature fluctuations in the P-NB state (Fig.~\ref{fig:thermal_characteristics}$a$) occur over a duration of several hundred seconds, with a characteristic time scale of about 350 s. These variations are presumably triggered by intermittent boiling events occurring during the P-NB state. While lacking techniques for temperature field measurements, we conducted Schlieren imaging that indirectly displays the local temperature gradient for several regimes. Although the Rayleigh number doesn't vary much in regime P-NB, the bubbles break the stability of the lower part of the thermal boundary layer (TBL) and trigger more plumes, resulting in enhanced thermal fluctuations (see also Movie S1). Shown in the green and orange outlined images (Fig.~\ref{fig:thermal_characteristics}$d$), the first few wakes and far-field wakes evidence the presence of large temperature differences between the wakes and the adjacent flow. This indirectly verifies the inhomogeneity of the flow in regime P-NB. However, in regime F-NB, fewer thermal plumes are present, corresponding to P-NB, indicating weaker thermal fluctuations. In addition, because no more wakes are visible in the images with orange outlines (from F-NB state in Fig.~\ref{fig:thermal_characteristics}$d$), the bubbly flow in the F-NB regime can be considered more homogeneous than that in the P-NB regime. However, even though the F-NB is homogeneous with relatively moderate thermal fluctuations, the key to the heat transfer increase lies in the increased frequency of the fluctuations. The characteristic \textcolor{black}{time scale} of the fluctuations is around {\color{black} 71s}. Although the fluctuations are moderate in amplitude, the high frequency in the F-NB state suggests the onset of turbulence in the system. 
While the latent heat of boiling and condensation do not contribute much to the total heat transfer, these are crucial for establishing a self-sustained biphasic cycle. The leading mechanism of heat transfer will be clear upon looking at the liquid velocity fluctuations.

Given that the latent heat contribution is less than 2$\%$ of the total heat flux, the mechanism of heat transfer enhancement and thermal intermittency reduction can be better appreciated from measurements of the local liquid velocity fluctuations both in the bulk and near the walls (see schematic in Fig.~\ref{fig:wall_velocity}$a$). {\color{black} The power spectral density (PSD) of the fluid velocity near the hot wall is shown in Fig.~\ref{fig:wall_velocity}$b$. In the NC state the kinetic energy of liquid velocity fluctuation is low. A clear scaling is missing as the NC flow is not turbulent.

From NC to P-NB to F-NB regime, the curves shift upward, indicative of a strong, monotonic increase in the overall kinetic energy of the velocity fluctuations. For the rising bubbles, we can define a characteristic frequency, $f_{c}\approx95~\text{Hz}$ \citep{mathai2020bubbly}. Interestingly, for frequencies higher than $f_c$, the kinetic energy spectra nearly follow a $-3$ scaling in both the P-NB and F-NB regimes and for both directions. The $-3$ scaling is indicative of dissipative scaling of energy and likely arises from the onset of the pseudo-turbulence due to bubbles induced agitation \citep{lance1991turbulence, prakash2016energy}. Once pseudo-turbulence is triggered, this mechanism dominates the heat transfer and results in the up to  246$\%$ enhancement of heat flux over the NC state.

In Fig.~\ref{fig:wall_velocity}$c$, we show a time series of liquid velocity fluctuations. Here again, we see that in regime NC, the velocity time series in the horizontal direction remains nearly constant. Next, once the HFE boils in regimes P-NB and F-NB, the increasing boiling results in stronger velocity fluctuations in both directions. In the horizontal direction, the velocity signals exhibit symmetry around $u^{'}_{\text{wall}}=0$. However, the fluctuations are asymmetric in the vertical direction. This behavior is characteristic of the pseudo-turbulence, wherein fluctuations are stronger in the positive (upward) direction. (see also Supplementary Material)

\begin{figure*}
\centering
\includegraphics[width=0.9\linewidth]{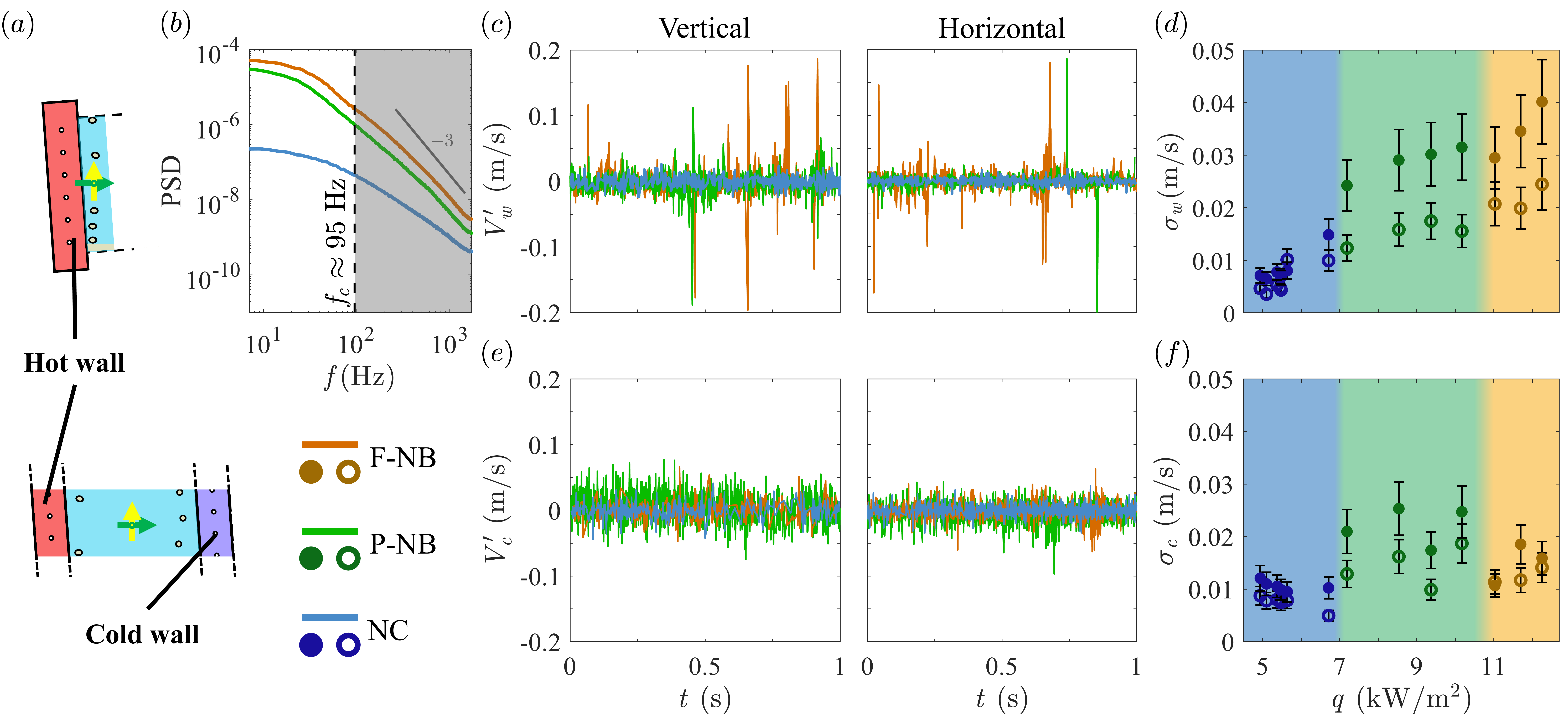}
\caption{Velocity fluctuations signals measured by Laser Doppler Anemometry (LDA). The upper row and the lower row are for the velocity signals near the hot wall and at the centroid (the bulk). Marked by the yellow dots at the crossing of the green horizontal and the yellow vertical arrows, ($a$) is the schematic of the velocity-measured positions where there is a 118.7 mm horizontal distance between two positions. ($b$) is the power density spectrum (PSD) of the velocity near the hot wall of the three regimes. $f_{c}\approx95\rm~Hz$ is the characteristics frequency. The second and third columns are the velocity fluctuations ($c$) near the hot wall and ($e$) at the centroid, which are for the velocity in the vertical and horizontal direction in the second and third columns. ($d$) and ($f$) are the standard deviation of velocity signals ($\sigma_{w}$ and $\sigma_{c}$) v.s. the wall heat flux near the hot wall and in the bulk.}
\label{fig:wall_velocity}
\end{figure*}

The standard deviation can also quantify the intensity of velocity fluctuations (see Fig.~\ref{fig:wall_velocity}$d$). In the horizontal direction, the standard deviation shows a gradual increase for all regimes, with only a marginal rise observed in regime P-NB, in contrast to a sudden increase for the vertical one. In the bulk region, the fluctuations of velocity are much weaker compared to those observed in the near-wall region, as illustrated in the second row of Fig.~\ref{fig:wall_velocity}$e$. The time series of horizontal velocity fluctuations shows only marginally higher fluctuations in the P-NB regime and more pronounced increments in the F-NB regime. The standard deviation of the velocity in the bulk is comparable to that of the near-wall region, as shown in Fig.~\ref{fig:wall_velocity}$f$, implying the extension of biphasic particles induced agitation at nearly all locations. The appearance of the extension in the F-NB regime implies a more thermally homogeneous convection, benefited from a strong mixing. This was also evident from the apparent invisibility of wakes in the shadowgraph images of regime F-NB (orange outlined images in Fig.~\ref{fig:thermal_characteristics}$d$) and from the low standard deviation of thermal fluctuations (Fig.~\ref{fig:thermal_characteristics}$c$).

\section{Conclusions}\label{sec:Summary}

In this work, we have demonstrated up to 246\% heat transfer enhancement in vertical convection by creating an {\it active biphasic} state of thermal convection. We show that a gas-liquid layer on the cell top enables nearly frictionless horizontal transport of the risen biphasic bubbles. By introducing a marginal tilt ($< 4^{\circ}$) to the VC cell, a gentle large-scale circulation (LSC) is generated that prevents the accumulation of the risen bubbles, worked on by the LSC.

We classify the Nusselt number $Nu$ measurements into three distinct regimes based on the boiling behavior and the thermal and liquid velocity fluctuations. Laser Doppler Anemometry (LDA) measurements reveal the onset of {\it pseudo-turbulence} in the liquid phase and the appearance of a $-3$ scaling for liquid kinetic energy spectra, characteristic of bubble-induced liquid agitation. This pseudo-turbulence drives the heat transfer increase with minor contribution from the latent heat. Beyond the onset of nucleate boiling (ONB), the heat transfer rises sharply with no observable change to the temperature of the hot wall. 

In summary, the present work has provided a novel strategy to enhance heat transfer that goes well beyond the limit of natural vertical convection. This system may be broadly applicable in closed natural convection scenarios requiring high heat transfer and intense local mixing without rigid moving parts, such as bioreactors or passive heat exchangers. Although we measured only up to 246\% increase in heat flux due to the heating elements reaching peak power capacity, in principle, the biphasic system has much higher up boundary for the heat transfer, which will be found in the future work.

\printcredits

\section*{Declaration of Competing Interest}

The authors declare that they have no known competing financial interests or personal relationships that could have appeared to influence the work reported in this paper.

\section*{Acknowledgments}
We thank Z. Wang and Y. Zhang for insightful discussions. This work is supported by NSFC Excellence Research Group Program for ‘Multiscale Problems in Nonlinear Mechanics’ (No. 12588201), the New Cornerstone Science Foundation through the New Cornerstone Investigator Program and the XPLORER PRIZE. 

\appendix
\section*{Supplementary material}

Supplementary data associated with this article can be found in a separate file. 

\bibliographystyle{plainnat} 
\bibliography{refer.bib}

\end{document}